\documentclass[10pt,prd,twocolumn,
nofootinbib,
noshowkeys,
noshowpacs,
superscriptaddress,
floatfix
]{revtex4-2}
\usepackage{amsmath,amsfonts,amsthm,amssymb}
\usepackage[dvips]{graphics,graphicx}
\usepackage[usenames,dvipsnames]{color}
\definecolor{darkblue}{RGB}{0,0,196}
\definecolor{darkgreen}{RGB}{0,120,0}
\usepackage[colorlinks=true,linktocpage=true,linkcolor=darkblue,citecolor=red,urlcolor=darkblue]{hyperref}
\usepackage{cancel}
\usepackage{bbold}
\usepackage{multirow}
\usepackage{longtable}
\usepackage{color}
\usepackage[normalem]{ulem}
\usepackage{hyperref}
\usepackage{bigints}
\usepackage{xparse}
\usepackage{physics}
\usepackage{verbatim}
\usepackage{minibox}
\usepackage{comment}
\usepackage{appendix}
\usepackage{slashed}
\usepackage{marginnote}
\usepackage{graphicx}
\usepackage[nice]{nicefrac}
\usepackage{amsmath}
\usepackage{hepunits}
\include{commands}
\usepackage{stackengine,scalerel}
\newcommand\hstar[1]{\ThisStyle{\ensurestackMath{%
  \setbox0=\hbox{$\SavedStyle#1$}%
  \stackengine{0pt}{\copy0}{\kern.2\ht0\smash{\SavedStyle\star}}{O}{c}{F}{T}{S}}}}

\definecolor {darkgreen}{rgb}{0.2,0.7,0.2}

\begin{document}
\title{Stability studies of first order spin-hydrodynamic frameworks}

\author{Asaad Daher}
\email{asaad.daher@ifj.edu.pl}
\affiliation{Institute  of  Nuclear  Physics  Polish  Academy  of  Sciences,  PL-31-342  Krak\'ow,  Poland}
\author{Arpan Das}
\email{arpan.das@ifj.edu.pl}
\affiliation{Institute  of  Nuclear  Physics  Polish  Academy  of  Sciences,  PL-31-342  Krak\'ow,  Poland}
\author{Radoslaw Ryblewski}
\email{radoslaw.ryblewski@ifj.edu.pl}
\affiliation{Institute  of  Nuclear  Physics  Polish  Academy  of  Sciences,  PL-31-342  Krak\'ow,  Poland}

\begin{abstract}
We study the stability of first-order dissipative spin-hydrodynamic frameworks.
We considered two different first-order dissipative spin-hydrodynamic frameworks.
The first one considers the spin chemical potential ($\omega^{\alpha\beta}$) to be first order ($\mathcal{O}(\partial)$) in the hydrodynamic gradient expansion. The hydrodynamic gradient ordering of the spin chemical potential is a debatable issue within the frameworks of spin hydrodynamics. Therefore as a second choice, we also consider the spin hydrodynamic equations with $\omega^{\alpha\beta}\sim\mathcal{O}(1)$. We find that for both frameworks, at the level of linear perturbations some spin modes can be unstable. To remove these generic instabilities we consider the \textit{Frenkel condition}. We argue that Frenkel condition helps get rid of the unstable solutions in both cases, but with a physical drawback for the case where $\omega^{\mu\nu}\sim\mathcal{O}(\partial)$.
\end{abstract}

\date{\today \hspace{0.2truecm}}

\maketitle
\flushbottom

\section{Introduction}
Experimental observations of spin-polarized weakly decaying hyperons have given us a unique opportunity to explore the vortical structure of the strongly coupled plasma produced in relativistic heavy-ion experiments~\cite{STAR:2017ckg, STAR:2018gyt,STAR:2019erd,ALICE:2019onw,ALICE:2019aid,STAR:2020xbm,Kornas:2020qzi,STAR:2021beb,ALICE:2021pzu,lisa2021}. Motivated by the successes of the relativistic dissipative hydrodynamic framework in heavy-ion phenomenology \cite{Florkowski:2017olj}, it is naturally desirable to generalize hydrodynamic framework to include spin as a dynamical degree of freedom. Several spin-hydrodynamic frameworks have been developed using relativistic kinetic theory~\cite{Florkowski:2017ruc,Florkowski:2017dyn,Florkowski:2018ahw,Florkowski:2018fap,Florkowski:2019qdp,Bhadury:2020puc,Bhadury:2020cop,Speranza:2020ilk,Weickgenannt:2020aaf,Weickgenannt:2021cuo,Shi:2020htn,Peng:2021ago,Sheng:2021kfc}, entropy current analysis~\cite{Hattori:2019lfp,Fukushima:2020ucl,Li:2020eon,She:2021lhe,Hongo:2021ona,Wang:2021ngp}, quantum statistical density operators~\cite{Becattini:2007nd,Becattini:2009wh,Becattini:2012pp,Becattini:2018duy,Hu:2021lnx}, effective Lagrangian approach~\cite{Montenegro:2017rbu,Montenegro:2017lvf,Montenegro:2018bcf,Montenegro:2020paq},  holography~\cite{Gallegos:2020otk,Garbiso:2020puw} and equilibrium partition functions~\cite{Gallegos:2021bzp}, etc.

Spin hydrodynamic frameworks are based on the conservation of the total angular momentum and the conservation of the total energy-momentum tensor.  The quantity that separates the spin-hydrodynamic framework from the standard hydrodynamic framework is the `spin chemical potential'. In the Navier-Stokes limit of the spin-hydrodynamic framework, the spin chemical potential is a hydrodynamic variable similar to the temperature, chemical potential, and fluid four-velocity. In local thermodynamic equilibrium, the hydrodynamic gradient ordering of the spin chemical potential is not a settled topic and different derivative counting schemes have been considered in literature~\cite{She:2021lhe,Hattori:2019lfp,Fukushima:2020ucl,Daher:2022xon,Wang:2021ngp,Hongo:2021ona}. Hydrodynamic gradient ordering of the spin chemical potential plays a crucial role in the thermodynamic as well as a hydrodynamic description with spin, e.g., if one considers that the spin chemical potential is $\mathcal{O}(1)$ or $\mathcal{O}(\partial^{})$ then this affects the first-order spin hydrodynamic description significantly.
Since the spin-hydrodynamic framework depends on the derivative ordering of the spin chemical potential then it is natural to investigate its effect on the propagation properties of linear perturbations.

Linear mode analysis of various hydrodynamic theories has been extensively discussed in the context of the stability and causality of fluid-dynamical theories. Generically one performs the stability and causality analysis around a hydrostatic state, where the spatial components of the flow velocity vanish. However, it should be pointed out that even if a theory is causal and stable in the hydrostatic limit, it does not necessarily imply stability in a boosted frame. Linear mode analysis of a generic first-order theory~\cite{Kovtun:2019hdm, Bemfica:2019knx, Bemfica:2017wps}, as well as second-order Israel-Stewart theory, have been scrutinized in various literature for a generic Lorentz boosted frame~\cite{Koide:2006ef,Denicol:2008ha,Romatschke:2009im,Pu:2009fj,Hiscock:1987zz,Hiscock:1985zz,Hiscock:1983zz}. Moreover using such studies it was found that causality and stability of various hydrodynamic models are intimately related. Note that in a hydrodynamic theory instability may appear due to various factors, but it has been explicitly shown that for a parameter space where the theory gives rise to acausal mode theory also show some instabilities~\cite{Pu:2009fj,Gavassino:2021kjm}. Such theoretical intricacies associated with the mode analysis has also been studied for the relativistic magnetohydrodynamics~\cite{Biswas:2020rps}, spin hydrodynamics~\cite{Ambrus:2022yzz,Hu:2022xjn,Hu:2022mvl,Sarwar:2022yzs}, chiral hydrodynamics~\cite{Speranza:2021bxf}, etc.

In the present work, we investigate the properties of various linear perturbations of the first-order dissipative spin-hydrodynamic framework. We show that the linear mode analysis crucially depends on the spin-hydrodynamic framework and the derivative ordering of the spin-chemical potential. In our analysis we consider two different spin hydrodynamic descriptions, one considers the spin chemical potential to be $\mathcal{O}(1)$ in the hydrodynamic gradient expansion~\cite{She:2021lhe} and the other theory considers the spin chemical potential to be $\mathcal{O}(\partial)$~\cite{Hattori:2019lfp,Fukushima:2020ucl,Hongo:2021ona,Daher:2022xon,Wang:2021ngp}. We also argue that both theories can give rise to linear modes which are unstable. Note that the spin chemical potential, denoted as $\omega^{\mu\nu}$, is a two-rank anti-symmetric tensor that has six independent components. Our calculations suggest that the instabilities appear in the spin-hydrodynamic description due to the boost degrees of freedom $\omega^{0i}$. In principle, such unstable modes can be removed from the theory by suitably eliminating $\omega^{0i}$ degrees of freedom. This can be achieved by incorporating the ``Frenkel condition"~\cite{Frenkel:1926zz, Cao:2022aku}.  

The paper is organized as follows. We begin by studying the stability of the first order spin-hydrodynamic \cite{Hattori:2019lfp}, which considers the spin chemical potential $\omega_{\mu\nu}$ to be of the first order in gradient expansion in Sec.~\ref{sec2}. In this section, we apply linear perturbation on top of a specified global equilibrium background for the spin-hydrodynamic equations, and then we solved them in Fourier space where we find that some spin modes can be unstable. We then show that imposing Frenkel condition gets rid of the unstable solutions while leading to some physical drawbacks to the system. In Sec.~\ref{sec3}, we repeat the same procedure but for the spin-hydrodynamic formulation \cite{She:2021lhe}, that follows the spin chemical potential at the leading order in the hydrodynamic gradient expansion. We also find that some spin modes are unstable. For such a case, imposing Frenkel condition gets rid of the unstable solution without any physical disadvantages. In Sec.~\ref{sec4} we summarize and conclude. 


\section{Spin chemical potential first order in the hydrodynamic gradient expansion}
\label{sec2}
\subsection{Formulation}
Often it has been argued that in the global equilibrium the spin chemical potential should be proportional to the thermal vorticity~\cite{Hattori:2019lfp,Weickgenannt:2020aaf}. Hence one can consider a situation where $\omega^{\alpha\beta}\sim\mathcal{O}(\partial)$. 
Such a spin-hydrodynamic description has been discussed in Refs.~\cite{Hattori:2019lfp,Fukushima:2020ucl,Daher:2022xon}. 
It will be similarly interesting to study the linear modes for the spin-hydrodynamic description where $\omega^{\alpha\beta}\sim\mathcal{O}(\partial)$. It should be emphasized that although the spin chemical potential is argued to be $\mathcal{O}(\partial)$, the spin density is $\mathcal{O}(1)$~\cite{Wang:2021ngp}. Considering that the spin density is proportional to the spin chemical potential 
brings non-triviality to this framework. We start with the energy-momentum tensor and spin tensor having the following form~\cite{Hattori:2019lfp,Fukushima:2020ucl,Daher:2022xon},
\begin{align}
        & T^{\mu\nu}=\varepsilon u^{\mu}u^{\nu}-p\Delta^{\mu\nu}+h^{\mu}u^{\nu}+h^{\nu}u^{\mu}\nonumber\\
        & ~~~~~~~+\tau^{\mu\nu}+q^{\mu}u^{\nu}-q^{\nu}u^{\mu}+\phi^{\mu\nu},\\
        & S^{\mu\alpha\beta}=u^{\mu} S^{\alpha \beta}+ S^{\mu\alpha\beta}_{(1)}.
\end{align}
We emphasize that in the above equation the energy-momentum tensor is not completely symmetric. Rather it contains an anti-symmetric part. Moreover, the spin tensor is only anti-symmetric in the last two indices. Such phenomenological energy-momentum tensor and spin tensor can be obtained from the canonical energy-momentum tensor and spin tensor using a proper pseudo-gauge transformation~\cite{Daher:2022xon}.
$h^{\mu}$ is the heat flow, $\tau^{\mu\nu}=\pi^{\mu\nu}+\Pi\Delta^{\mu\nu}$ is the dissipative corrections to the symmetric part of the energy-momentum tensor.
$\pi^{\mu\nu}$ is the traceless part of $\tau^{\mu\nu}$ and it is related to the shear-viscosity ($\eta$). On the other hand, $\Pi$ is related to bulk viscosity ($\zeta$). The dissipative corrections to the anti-symmetric part of the energy-momentum tensor are $q^{\mu}$ and $\phi^{\mu\nu}$. $h^{\mu}$, $\tau^{\mu\nu}$, $q^{\mu}$ and $\phi^{\mu\nu}$ satisfy following conditions, $h^{\mu}u_{\mu}=0$,
$\tau^{\mu\nu}=\tau^{\nu\mu}$, $\tau^{\mu\nu}u_{\nu}=0$, $q^{\mu}u_{\mu}=0$, 
$\phi^{\mu\nu}=-\phi^{\nu\mu}$ and $\phi^{\mu\nu}u_{\nu}=0$. The dissipative correction to the spin tensor, i.e. $S^{\lambda\mu\nu}_{(1)}$ is not fixed at the level of first order dissipative spin hydrodynamics as it does not contribute to the non-equilibrium entropy current\footnote{Recall that the ansatz for the non-equilibrium entropy current can be expressed as $\mathcal{S}^{\mu}=\beta_{\nu} T^{\mu \nu}+p \beta^{\mu}-\beta \omega_{\alpha \beta} S^{\mu \alpha \beta}$. For the first order dissipative spin hydrodynamic framework,  $\mathcal{S}^{\mu}$ can contain terms up to order $\mathcal{O}(\partial)$. Such $\mathcal{O}(\partial)$ terms can come from both $T^{\mu\nu}$ and $S^{\mu\alpha\beta}$. But if we consider $\omega^{\alpha\beta}$ to be of the order $\mathcal{O}(\partial)$, then $\omega_{\alpha\beta}S^{\mu\alpha\beta}_{(1)}$ terms will be of the order $\mathcal{O}(\partial^2)$. Therefore we can conclude that if $\omega^{\alpha\beta}$ is of the order $\mathcal{O}(\partial)$ then the dissipative part of the spin tensor would not contribute to the entropy current. On the other hand, if  $\omega^{\alpha\beta}$ is of the order $\mathcal{O}(1)$ then $S^{\mu\alpha\beta}_{(1)}$ will contribute to the first order spin-hydrodynamic framework. This is the most striking difference between various frameworks considered here.}~\cite{Hattori:2019lfp,Fukushima:2020ucl,Daher:2022xon}. Various dissipative currents can be uniquely determined by using the condition that for an isolated dissipative system entropy will be produced. In terms of the hydrodynamic variables, i.e., $T$, $u^{\mu}$, and $\omega^{\mu\nu}$ it can be shown that~\cite{Daher:2022xon,Fukushima:2020ucl,Hattori:2019lfp}, 
\begin{align}
    & h^{\mu}  =-\kappa\bigg(Du^{\mu}-\beta\nabla^{\mu}T\bigg)\label{equ3ver2}\\
    & q^{\mu}  =\lambda\bigg(Du^{\mu}+\beta\nabla^{\mu}T-4\omega^{\mu\nu}u_{\nu}\bigg)\label{equ4ver2}\\
    & \tau^{\mu\nu}= 
    \eta\bigg(\Delta^{\mu\alpha}\partial_{\alpha}u^{\nu}+\Delta^{\nu\alpha}\partial_{\alpha}u^{\mu}-\frac{2}{3}\Delta^{\mu\nu}\Delta^{\alpha\beta}\partial_{\beta}u_{\alpha}\bigg)\nonumber\\
    & ~~~~~~~~~~~~~~~~~~~~~~~~~~~~~~~~~~~~~~~~~~+\zeta (\partial_{\alpha}u^{\alpha}) \Delta^{\mu\nu}\label{equ5ver2}\\
    & \phi^{\mu\nu} 
    = \widetilde{\gamma}\bigg(\nabla^{\mu}u^{\nu}-\nabla^{\nu}u^{\mu}+4\Delta^{\mu\alpha}\Delta^{\nu\beta}\omega_{\alpha\beta}\bigg).\label{equ6ver2}
\end{align}
Here $D\equiv u^{\mu}\partial_{\mu}$, $\nabla^{\mu}=\Delta^{\mu\nu}\partial_{\nu}$, and $\widetilde{\gamma}=\beta\gamma/2$. Note that $h^{\mu}$, $q^{\mu}$, $\tau^{\mu\nu}$ and $\phi^{\mu\nu}$ are $\mathcal{O}(\partial)$ in the hydrodynamic gradient expansion. Dissipative currents $h^{\mu}$ and $q^{\mu}$ as given in Eqs.~\eqref{equ3ver2} and ~\eqref{equ4ver2} can be further simplified by using the leading order hydrodynamic equations. Conservation of $T^{\mu\nu}_{(0)}\equiv \varepsilon u^{\mu}u^{\nu}-p\Delta^{\mu\nu}$ gives us, 
\begin{align}
    & u^{\mu}\partial_{\mu}\varepsilon+(\varepsilon+p)(\partial_{\mu}u^{\mu})=0\\
    & (\varepsilon+p)Du^{\alpha}-\nabla^{\alpha}p=0.\label{equ9ver2}
\end{align}
Using Eq.~\eqref{equ9ver2} back into Eq.~\eqref{equ3ver2} it can be easily shown that,  
\begin{align}
    h^{\mu} & = 0+\mathcal{O}(\partial^2).
\end{align}
In order to obtain the above equation we have used the thermodynamic relation $Ts+S^{\alpha\beta}\omega_{\alpha\beta}=\varepsilon+p$, $dp=sdT+S^{\mu\nu}d\omega_{\mu\nu}$. 
Here we consider $\omega^{\mu\nu}\sim\mathcal{O}(\partial)$, and $S^{\mu\nu}\sim\mathcal{O}(1)$ in the hydrodynamic gradient expansion~\cite{Hattori:2019lfp,Fukushima:2020ucl,Daher:2022xon}. Furthermore, 
\begin{align}
    q^{\mu} & =\lambda\bigg(Du^{\mu}+\beta\nabla^{\mu}T-4\omega^{\mu\nu}u_{\nu}\bigg)\nonumber\\
     & = \lambda\bigg(2\frac{\nabla^{\mu}p}{\varepsilon+p}-4\omega^{\mu\nu}u_{\nu}\bigg)+\mathcal{O}(\partial^2).
\end{align}
To obtain the linear order hydrodynamic perturbation with respect to a global equilibrium we consider, $u^{\mu}_{(0)}\equiv(1,0,0,0)$, $\omega^{\mu\nu}_{(0)}=0$, and $S^{\mu\nu}_{(0)}=0$~\cite{Hattori:2019lfp}. For such a global equilibrium configuration various dissipative currents, i.e., $q^{\mu}_{(0)}=0$, $\phi^{\mu\nu}_{(0)}=0$ and $\tau^{\mu\nu}_{(0)}=0$. Note $\tau^{\mu\nu}$, $q^{\mu}$ and $\phi^{\mu\nu}$ are already $\mathcal{O}(\partial)$. Therefore we consider $\delta\tau^{\mu\nu}$, $\delta q^{\mu}$ and $\delta \phi^{\mu\nu}$ up to $\mathcal{O}(\partial^2)$ and we neglect all higher order terms,  
\begin{align}
    & \delta\tau^{\mu\nu}= \eta\bigg(\Delta^{\mu\alpha}_{(0)}\partial_{\alpha}\delta u^{\nu}+\Delta^{\nu\alpha}_{(0)}\partial_{\alpha}\delta u^{\mu}-\frac{2}{3}\Delta^{\mu\nu}_{(0)}\Delta^{\alpha\beta}_{(0)}\partial_{\beta}\delta u_{\alpha}\bigg)\nonumber\\
    & ~~~~~~~~~~~~~+\zeta (\partial_{\alpha}\delta u^{\alpha}) \Delta^{\mu\nu}_{(0)}+\mathcal{O}(\partial^3).
    \label{equ7ver1}\\
    &  \delta q^{\mu}  
      = \lambda\bigg(2\frac{\Delta^{\mu\alpha}_{(0)}\partial_{\alpha}\delta p}{\varepsilon_{(0)}+p_{(0)}}-4\delta \omega^{\mu\nu}u_{\nu}^{(0)}\bigg)+\mathcal{O}(\partial^3).
     \label{equ9ver1}\\
    &  \delta\phi^{\mu\nu} 
     = \widetilde{\gamma}\bigg(\Delta^{\mu\alpha}_{(0)}\partial_{\alpha}\delta u^{\nu}-\Delta^{\nu\alpha}_{(0)}\partial_{\alpha}\delta u^{\mu}\nonumber\\
     & ~~~~~~~~~~~~~~~~~~~~~+4\Delta^{\mu}_{~\rho(0)}\Delta^{\nu}_{~\lambda(0)}\delta \omega^{\rho\lambda}\bigg)+\mathcal{O}(\partial^3).
\label{equ11ver1}
\end{align}
In global equilibrium, it is easy to show that, $T^{0i}_{(0)}=0$. The perturbation $\delta T^{0i}$ can be expressed as,
\begin{align}
    \delta T^{0i}
    & = (\varepsilon_{(0)}+p_{(0)})\delta u^i+\delta\tau^{0i}-\delta q^i+\delta \phi^{0i}.
    \label{equ13ver1}
\end{align}
For the flow perturbation of the form, $\delta u^{\mu}=(0,\delta u^i)$, using Eqs.~\eqref{equ7ver1}-\eqref{equ11ver1} it can be shown that $\delta \tau^{0i}=0$, $\delta q^{0}=0$, and $\delta \phi^{0i}=0$. But $\delta q^i$ is non vanishing and it can be expressed as, 
  \begin{align}
    \delta q^{i}  & = \lambda\bigg(2\frac{\nabla^{i}_{(0)}\delta p}{\varepsilon_{(0)}+p_{(0)}}-4\delta \omega^{i\nu}u_{\nu}^{(0)}\bigg)+\mathcal{O}(\partial^3)\nonumber\\
     & = \lambda^{\prime}c_s^2\partial^i\delta\varepsilon-\frac{4\lambda}{\chi_b}\delta S^{i0}+\mathcal{O}(\partial^3)\nonumber\\
     & = \lambda^{\prime}c_s^2\partial^i\delta\varepsilon-D_b\delta S^{i0}+\mathcal{O}(\partial^3).
     \label{equ15ver1}
\end{align}
Here we have defined $\lambda^{'}=\frac{2\lambda}{\varepsilon_{(0)}+p_{(0)}}$, $c_s^2=\partial p/\partial\varepsilon$,  $D_b=4\lambda/\chi_b$, and $\chi_b=\partial S^{i0}/\partial \omega^{i0}$. 
We consider $c_s^2$, $\chi_b$, and $D_b$ to be constants as any space-time derivative of these quantities will give rise to higher order terms. 
Therefore the perturbation $\delta T^{0i}\equiv\delta\pi^i$ can be expressed as,
\begin{align}
      \delta\mathfrak{\pi}^i & =(\varepsilon_{(0)}+p_{(0)})\delta u^i\nonumber\\
      & ~~~~~~~~~~~-\lambda^{\prime}c_s^2\partial^i\delta\varepsilon+D_b\delta S^{i0}+\mathcal{O}(\partial^3).
      \label{equ16ver2}
\end{align}
Note $\delta\pi^i$ contains terms of the order of $\mathcal{O}(\partial)$ and higher. Moreover, $\lambda^{\prime}$ and $D_b$ originate from the anti-symmetric part of the energy-momentum tensor. 
Conservation of the total angular momentum can be used to write the evolution equation for the spin tensor, 
\begin{align}
 & u^{\mu}\partial_{\mu}S^{\alpha\beta}+S^{\alpha\beta}\partial_{\mu}u^{\mu}\nonumber\\
& ~~~~~~~~~~=-2\bigg(q^{\alpha}u^{\beta}-q^{\beta}u^{\alpha}+\phi^{\alpha\beta}\bigg).
\end{align}
Noting that we are considering the global equilibrium with $S^{\mu\nu}_{(0)}=0$ and $\omega^{\mu\nu}_{(0)}=0$, at the level of linear order perturbation we can write, 
\begin{align}
& \partial_{0}\delta S^{\alpha\beta}= -2\bigg(\delta q^{\alpha}u^{\beta}_{(0)}-\delta q^{\beta}u^{\alpha}_{(0)}+\delta \phi^{\alpha\beta}\bigg)+\mathcal{O}(\partial^3).
\label{equ19ver1}
\end{align}
Using the above equation we can obtain the evolution equation for $\delta S^{0i}$ and $\delta S^{ij}$. The evolution equation of $\delta S^{0i}$ is, 
\begin{align}
    \partial_{0}\delta S^{0i}& = -2\bigg(\delta q^{0}u^{i}_{(0)}-\delta q^{i}u^{0}_{(0)}+\delta \phi^{0i}\bigg)+\mathcal{O}(\partial^3)\nonumber\\
    & = 2~\delta q^i +\mathcal{O}(\partial^3)\nonumber\\
    & = 2\lambda^{\prime}c_s^2\partial^i\delta\varepsilon-2D_b\delta S^{i0}+\mathcal{O}(\partial^3). 
    \label{equ22ver2}
    \end{align}
    Using Eq.~\eqref{equ19ver1} the evolution equation of $\delta S^{ij}$ can be written as, 
    \begin{align}
        & \partial_0\delta S^{ij} = -2\delta\phi^{ij}+\mathcal{O}(\partial^3)\nonumber\\
        & = -2D_s\delta S^{ij}-2\gamma^{\prime}(\partial^i\delta\pi^j-\partial^j\delta\pi^i)+\mathcal{O}(\partial^3).
        \label{equ23ver2}
    \end{align}
    Here $D_s=4\widetilde{\gamma}/\chi_s$ and $\chi_s=\partial S^{ij}/\partial \omega^{ij}$. 
    The longitudinal projection of the conservation of the total energy-momentum tensor, $u_{\nu}\partial_{\mu}T^{\mu\nu}=0$ implies,
\begin{align}
& u^{\mu}\partial_{\mu}\varepsilon+(\varepsilon+p)\partial_{\mu}u^{\mu}=-u_{\nu}\partial_{\mu}T^{\mu\nu}_{(1)}
\label{equ23ver1}
\end{align}
Note that the L.H.S of the above equation is $\mathcal{O}(\partial)$, but the R.H.S is $\mathcal{O}(\partial^2)$. Therefore for the perturbation equation, L.H.S will be up to order $\mathcal{O}(\partial^2)$ and the R.H.S will be up to $\mathcal{O}(\partial^3)$. Such a perturbation equation can be expressed as, 
\begin{align}
  & \partial_{0}\delta\varepsilon+(\varepsilon_{(0)}+p_{(0)})\partial_{i}\delta u^{i}\nonumber\\
        & ~~~~~= -\partial_0\delta\tau^{00}-\partial_{i}\bigg[\delta\tau^{i0}+\delta q^{i}+\delta\phi^{i0}\bigg].
        \label{equ24ver1}
\end{align}
In order to obtain Eq.~\eqref{equ24ver1} we drop some terms, e.g. $\delta u^{\mu}\partial_{\mu}\delta\varepsilon$ and $(\delta\varepsilon+\delta p)\partial_{\mu}\delta u^{\mu}$ which are $\mathcal{O}(\partial^3)$. 
Since we are only restricting our analysis for linear modes we have dropped terms which are non-linear in perturbations. Using the conditions that $\delta\tau^{00}=0$, $\delta\tau^{0i}=0$, $\delta\phi^{i0}=0$, and the expression of $\delta q^i$ back into Eq.~\eqref{equ24ver1} it can be shown that,
\begin{align}
& \partial_0\delta\varepsilon+\partial_i\delta\pi^i+2\bigg(\lambda^{\prime}c_s^2\partial_i\partial^i\delta\varepsilon-D_b\partial_i\delta S^{i0}\bigg)=0.
    \label{equ28ver2}
    \end{align}
    In the above equation, we have not considered $\mathcal{O}(\partial^4)$ terms. Taking the normal projection of total energy-momentum tensor, $\Delta^{\alpha}_{~\nu}\partial_{\mu}T^{\mu\nu}=0$ we find, 
\begin{align}
&  (\varepsilon+p)Du^{\alpha}-\Delta^{\alpha\beta}\partial_{\beta}p+\Delta^{\alpha}_{~\nu}\partial_{\mu}\tau^{\mu\nu}+\Delta^{\alpha}_{~\nu}\partial_{\mu}(q^{\mu}u^{\nu})\nonumber\\
& ~~~~~~~~-\Delta^{\alpha}_{~\nu}\partial_{\mu}(q^{\nu}u^{\mu})+\Delta^{\alpha}_{~\nu}\partial_{\mu}\phi^{\mu\nu}=0. 
\label{equ28ver1}
\end{align}
We should emphasize that Eq.~\eqref{equ28ver1} contains term up to $\mathcal{O}(\partial^2)$. Therefore the linear order perturbation of Eq.~\eqref{equ28ver1} must contain terms up to order $\mathcal{O}(\partial^3)$ and we can neglect higher order terms.  The perturbation equation associated with Eq.~\eqref{equ28ver1} can be expressed as, 
\begin{align}
    & (\varepsilon_{(0)}+p_{(0)})\partial_0\delta u^{\alpha}-\Delta^{\alpha\beta}_{(0)}\partial_{\beta}\delta p\nonumber\\
    & +\eta\Delta^{\alpha}_{~\nu(0)}\Delta^{\mu\beta}_{(0)}\partial_{\mu}\partial_{\beta}\delta u^{\nu}+\eta \Delta^{\alpha\beta}_{(0)}\partial_{\mu}\partial_{\beta}\delta u^{\mu}\nonumber\\
    & -\frac{2}{3}\eta\Delta^{\alpha\mu}_{(0)}\partial_{\mu}\partial_{\delta}\delta u^{\delta}+\zeta \Delta^{\alpha\mu}_{(0)}\partial_{\mu}\partial_{\delta}\delta u^{\delta}\nonumber\\
    & -\Delta^{\alpha}_{~\nu(0)} \partial_0\delta q^{\nu}+\Delta^{\alpha}_{~\nu(0)}\partial_{\mu}\delta\phi^{\mu\nu}=0.
\end{align}
For $\alpha=0$ the L.H.S of the above equation identically vanishes. For $\alpha=i$ we find, 
\begin{align}
    & \partial_0\delta\pi^i-c_s^2\partial^i\delta\varepsilon+(\gamma_{\perp}+\gamma^{\prime})(\delta^i_{~j}\partial^{k}\partial_k-\partial^i\partial_j)\delta\pi^j\nonumber\\
    & ~~~~~~~~~~~~+\gamma_{||}\partial^i\partial_k\delta\pi^k+D_s\partial_k\delta S^{ki}=0.
    \label{equ31ver4}
\end{align}
Throughout the derivation we use the following notations, 
\begin{align}
    & c_s^2\equiv \frac{\partial p}{\partial\varepsilon},~~ \chi_s\equiv \frac{\partial S^{ij}}{\partial\omega^{ij}},~~D_s\equiv \frac{4\widetilde{\gamma}}{\chi_s},~~\gamma^{\prime}\equiv\frac{\widetilde{\gamma}}{\varepsilon_{(0)}+p_{(0)}},\nonumber\\
    &\chi_b\equiv\frac{\partial S^{i0}}{\partial\omega^{i0}}, ~~D_b\equiv \frac{4\lambda}{\chi_b},~~\lambda^{\prime}\equiv\frac{2\lambda}{\varepsilon_{(0)}+p_{(0)}}\nonumber\\
    & \gamma_{||}\equiv\frac{1}{\varepsilon_{(0)}+p_{(0)}}(\zeta+\frac{4}{3}\eta),~~\gamma_{\perp}\equiv \frac{\eta}{\varepsilon_{(0)}+p_{(0)}}.
    \label{equ26new}
\end{align}
Eqs.~\eqref{equ22ver2}, \eqref{equ23ver2}, \eqref{equ28ver2} and \eqref{equ31ver4} along with Eq.~\eqref{equ16ver2} are main perturbation equations. We emphasize that for this framework the standard fluid perturbations, i.e., $\delta\varepsilon$, $\delta u^i$ are coupled with the spin perturbation $\delta S^{0i}$, and $\delta S^{ij}$.

\subsection{Fourier space equations}
Eqs.~\eqref{equ22ver2}, \eqref{equ23ver2}, \eqref{equ28ver2} and \eqref{equ31ver4} can be solved in the momentum space to obtain different dispersion relations associated with different perturbation modes. Various perturbations, $\delta\varepsilon$, $\delta\pi^k$, $\delta S^{ij}$ and $\delta S^{0i}$ can be expressed as plane waves in the following way 
\begin{align}
    & \delta\varepsilon=\widetilde{\delta\varepsilon}~e^{-i\omega t+i\vec{k}\cdot\vec{x}}\nonumber\\
    & \delta\pi^k=\widetilde{\delta\pi^k}~e^{-i\omega t+i\vec{k}\cdot\vec{x}}\nonumber\\
    & \delta S^{ij}=\widetilde{\delta S^{ij}}~e^{-i\omega t+i\vec{k}\cdot\vec{x}}\nonumber\\
    & \delta S^{0i}=\widetilde{\delta S^{0i}}~e^{-i\omega t+i\vec{k}\cdot\vec{x}}.
\end{align}
Due to the rotational symmetry of the system, we can consider waves which are propagating only along the $z$ direction, i.e. $\vec{k}=(0,0,k_z)$. For such a choice of plane wave Eqs.~\eqref{equ22ver2}, \eqref{equ23ver2}, \eqref{equ28ver2} and \eqref{equ31ver4} become, 
\begin{align}
& -i\omega\widetilde{\delta\varepsilon}+ik_z\widetilde{\delta\pi^z}+2\lambda^{\prime}c_s^2k_z^2\widetilde{\delta\varepsilon}+2D_bik_z\widetilde{\delta S^{0z}}=0\label{equ30ver2}\\
& -i\omega\widetilde{\delta\pi^x}+(\gamma_{\perp}+\gamma^{\prime})k_z^2\widetilde{\delta\pi^x}+ik_zD_s\widetilde{\delta S^{zx}}=0\label{equ31ver2}\\
& -i\omega\widetilde{\delta\pi^y}+(\gamma_{\perp}+\gamma^{\prime})k_z^2\widetilde{\delta\pi^y}+ik_zD_s\widetilde{\delta S^{zy}}=0\label{equ32ver2}\\
& -i\omega\widetilde{\delta\pi^z}+ik_zc_s^2\widetilde{\delta\varepsilon}+\gamma_{||}k_z^2\widetilde{\delta\pi^z}=0\label{equ33ver2}\\
& -i\omega\widetilde{\delta S^{xy}}+2D_s\widetilde{\delta S^{xy}}=0\label{equ34ver2}\\
& -i\omega\widetilde{\delta S^{zx}}+2D_s \widetilde{\delta S^{zx}}-2\gamma^{\prime}(ik_z)\widetilde{\delta \pi^x}=0\label{equ35ver2}\\
& -i\omega\widetilde{\delta S^{yz}}+2D_s \widetilde{\delta S^{yz}}+2\gamma^{\prime}(ik_z)\widetilde{\delta \pi^y}=0\label{equ36ver2}\\
& -i\omega\widetilde{\delta S^{0x}}-2D_b\widetilde{\delta S^{0x}}=0\label{equ37ver2}\\
& -i\omega\widetilde{\delta S^{0y}}-2D_b\widetilde{\delta S^{0y}}=0\label{equ38ver2}\\
& -i\omega\widetilde{\delta S^{0z}}+2\lambda^{\prime}c_s^2(ik_z)\widetilde{\delta\varepsilon}-2D_b\widetilde{\delta S^{0z}}=0.\label{equ39ver2}
\end{align}
These equations can be represented as a matrix equation $M\times v=0$ such that:
\begin{align*}
v=\bigg(\widetilde{\delta\varepsilon},\widetilde{\delta\pi^z},\widetilde{\delta S^{0z}},\widetilde{\delta\pi^x},\widetilde{\delta S^{zx}},\widetilde{\delta\pi^y},\widetilde{\delta S^{zy}},\widetilde{\delta S^{0x}},\widetilde{\delta S^{0y}},\widetilde{\delta S^{xy}}\bigg)^T
\end{align*}
For such a choice, $\widetilde{\delta\varepsilon},\widetilde{\delta\pi^z},\widetilde{\delta S^{0z}}$ can be grouped together to write,   
\begin{align}
 &   \begin{pmatrix}
-i\omega+2\lambda^{\prime}c_s^2k_z^2 & ik_z & 2D_bik_z\\
ic_s^2k_z & -i\omega+\gamma_{||}k_z^2 & 0\\
2\lambda^{\prime}c_s^2ik_z & 0 & -i\omega-2D_b.
\end{pmatrix}
\begin{pmatrix}
\widetilde{\delta\varepsilon} \\
\widetilde{\delta\pi^z} \\
\widetilde{\delta S^{0z}}
\end{pmatrix}\nonumber\\
&~~~~~~~~~~~~~~~~~~~~~~~~~~~~~~~~~~~~\equiv\mathcal{A}\begin{pmatrix}
\widetilde{\delta\varepsilon} \\
\widetilde{\delta\pi^z} \\
\widetilde{\delta S^{0z}}
\end{pmatrix}.
\end{align}
Similarly $\widetilde{\delta\pi^x},\widetilde{\delta S^{zx}}$ and $\widetilde{\delta\pi^y},\widetilde{\delta S^{zy}}$ can be grouped together in the following way, 
\begin{align}
 &   \begin{pmatrix}
-i\omega+(\gamma_{\perp}+\gamma^{\prime})k_z^2 & ik_zD_s\\
-2\gamma^{\prime}ik_z & -i\omega+2D_s 
\end{pmatrix}
\begin{pmatrix}
\widetilde{\delta\pi^x} \\
\widetilde{\delta S^{zx}}
\end{pmatrix}\nonumber\\
& ~~~~~~~~~~~~~~~~~~~~~~~~~~~~~~~~~~~~~~~\equiv\mathcal{B}\begin{pmatrix}
\widetilde{\delta\pi^x} \\
\widetilde{\delta S^{zx}}
\end{pmatrix}
\label{equ40ver2}
\end{align}
and,
\begin{align}
  &  \begin{pmatrix}
-i\omega+(\gamma_{\perp}+\gamma^{\prime})k_z^2 & ik_zD_s\\
-2\gamma^{\prime}ik_z & -i\omega+2D_s 
\end{pmatrix}
\begin{pmatrix}
\widetilde{\delta\pi^y} \\
\widetilde{\delta S^{zy}}
\end{pmatrix}\nonumber\\
& ~~~~~~~~~~~~~~~~~~~~~~~~~~~~~~~~~~~~~~~\equiv\mathcal{B}
\begin{pmatrix}
\widetilde{\delta\pi^y} \\
\widetilde{\delta S^{zy}}
\end{pmatrix}.
\label{equ41ver2}
\end{align}
The spin parts $\widetilde{\delta S^{0x}},\widetilde{\delta S^{0y}}$, and $\widetilde{\delta S^{xy}}$ are not coupled to other perturbations. Therefore Eq.~\eqref{equ34ver2}, Eq.~\eqref{equ37ver2} and Eq.~\eqref{equ38ver2} can be written as, 
\begin{align}
& \begin{pmatrix}
-i\omega-2D_b & 0 & 0\\
0 & -i\omega-2D_b & 0\\
0 & 0 & -i\omega+2D_s
\end{pmatrix}
\begin{pmatrix}
\widetilde{\delta S^{0x}}\\
\widetilde{\delta S^{0y}}\\
\widetilde{\delta S^{xy}}
\end{pmatrix}\nonumber\\
&~~~~~~~~~~~~~~~~~~~~~~~~~~~~~~~~~~~~~~~~~~~~~~\equiv\mathcal{C}\begin{pmatrix}
\widetilde{\delta S^{0x}}\\
\widetilde{\delta S^{0y}}\\
\widetilde{\delta S^{xy}}
\end{pmatrix}.
\end{align}
Various dispersion relations can be obtained by solving $det M=0$, where 
\begin{align}
   M= \begin{pmatrix}
    \mathcal{A}&0_{3\times2}&0_{3\times2}&0_{3\times3}\\
    0_{2\times3}&\mathcal{B}&0_{2\times2}&0_{2\times3}\\
    0_{2\times3}&0_{2\times2}&\mathcal{B}&0_{2\times3}\\
    0_{3\times3}&0_{3\times2}&0_{3\times2}&\mathcal{C}
    \end{pmatrix}
\end{align}
Using the advantage of $M$ being a block diagonal matrix, we can directly obtain the determinant such that  
\begin{align}
    \text{det}(\mathcal{A})\text{det}(\mathcal{B})^2(-i\omega-2D_b)^2(-i\omega+2D_s)=0,
\end{align}
and can be summarised as, 
\begin{align}
    &\omega=-2iD_{s}\\
    &\omega=+2iD_{b}\label{wrong1} \quad (\text{two modes})\\
    &\omega=-2iD_{s}-i\gamma^{\prime}k_{z}^{2}+\mathcal{O}(k_{z}^{4})~\quad (\text{two modes})\\
    &\omega=-i\gamma_{\perp}k_{z}^{2}+\mathcal{O}(k_{z}^{4}) \quad(\text{two modes})\\
    &\omega=+c_{s}k_{z}-\frac{i}{2}\gamma_{\parallel}k_{z}^{2}+\mathcal{O}(k_{z}^{3})\\
    &\omega=-c_{s}k_{z}-\frac{i}{2}\gamma_{\parallel}k_{z}^{2}+\mathcal{O}(k_{z}^{3})\\
    &\omega=2iD_{b}-2ic_{s}^{2}\lambda^{'}k_{z}^{2}+\mathcal{O}(k_{z}^{4})\label{wrong2}.
\end{align}
Note that a physical plane wave of the form $e^{-i\omega t+i\vec{k}\cdot\vec{x}}$, must not give a solution which is growing with time. But from Eq.~\eqref{wrong1} and \eqref{wrong2} it is clear that unstable mode exists for $D_b>0$. Therefore the spin hydrodynamic equations are not stable for the linear perturbation around the global equilibrium considered here. The same observation has been pointed out in Ref.~\cite{Sarwar:2022yzs}~\footnote{Note that in Ref.~\cite{Sarwar:2022yzs} authors considered a different counting scheme than what we considered here. In this reference, the authors could find next to the leading order contribution to the spin tensor and the associated transport coefficient $\chi_1$. In the limit $\chi_1=0$, various dissipative current as obtained in Ref.~\cite{Sarwar:2022yzs} matches with the calculation as given in Ref.~\cite{Hattori:2019lfp}. In this limit there exists an unstable mode if $D_b>0$ or $\chi_b>0$.}. This problem may be solved by appropriately defining $D_b$. There could be a physical reason to choose an appropriate sign of $D_b$\footnote{Throughout the calculation, $D_b$ and $D_s$ are considered to be positive. This implies that $\chi_b>0$ and $\chi_s>0$. This is very natural to argue using the equation of state relating spin density tensor $S^{\mu\nu}$ and $\omega^{\mu\nu}$~\cite{Wang:2021ngp}. Generically the spin density tensor should be proportional to the spin chemical potential, i.e. $S^{\mu\nu}\sim \omega^{\mu\nu}$. The proportionality factor should either be positive or negative. If the proportionality factor is negative then $D_b<0$ and $D_s<0$. In this case, although $D_b<0$, due to negative $D_s$ some modes will remain unstable. On the other hand, if the proportionality factor is positive then $D_b$ and $D_s$ are both positive. In that case, also the theory will give rise to unstable modes. Only if $D_b<0$ or $\chi_b<0$, keeping $\chi_s>0$, then unstable modes will not appear. This implies that the proportionality factor appearing in the spin equation of state for the $0i$-th components will be different from $ij$-th components. This is rather difficult to justify physically.}. But this needs to be investigated thoroughly for a proper understanding of the spin perturbation equations. We consider an alternative approach to remove this instability by imposing the Frenkel condition. We know that this instability is caused by the modes associated with $D_b$, and it may be possible to eliminate those problematic modes using the Frenkel condition, i.e. $S^{\mu\nu}u_{\nu}=0=\omega^{\mu\nu}u_{\nu}$~\cite{Cao:2022aku}.


\subsection{Frenkel condition to solve the problem of instability}
If we impose the Frenkel condition, i.e. $S^{\mu\nu}u_{\nu}=0$ or $\omega^{\mu\nu}u_{\nu}=0$, then various dissipative currents, i.e. $h^{\mu}$, $\tau^{\mu\nu}$ as given in Eqs.~\eqref{equ3ver2},\eqref{equ5ver2} respectively remain unaltered. But $q^{\mu}$ and $\phi^{\mu\nu}$ change to, 
\begin{align}
    & q^{\mu}  =\lambda\bigg(Du^{\mu}+\beta\nabla^{\mu}T\bigg).\\
    & \phi^{\mu\nu} = \widetilde{\gamma}\bigg(\nabla^{\mu}u^{\nu}-\nabla^{\nu}u^{\mu}+4\omega^{\mu\nu}\bigg).
\end{align}
Neglecting all higher order terms,  $\delta q^{\mu}$ and $\delta\phi^{\mu\nu}$ up to $\mathcal{O}(\partial^2)$ can be expressed as,
\begin{align}
    & \delta q^{\mu}  
      = \lambda\bigg(2\frac{\Delta^{\mu\alpha}_{(0)}\partial_{\alpha}\delta p}{\varepsilon_{(0)}+p_{(0)}}\bigg)+\mathcal{O}(\partial^3).\label{equ53ver2}\\
     & \delta\phi^{\mu\nu} = \widetilde{\gamma}\bigg(\Delta^{\mu\alpha}_{(0)}\partial_{\alpha}\delta u^{\nu}-\Delta^{\nu\alpha}_{(0)}\partial_{\alpha}\delta u^{\mu}+4\delta \omega^{\mu\nu}\bigg)+\mathcal{O}(\partial^3).
     \label{equ54ver2}
\end{align}
Using the evolution equation of $\delta S^{\alpha\beta}$ as given in Eq.~\eqref{equ19ver1} we find, 
\begin{align}
    \partial_{0}\delta S^{0i}
    &= 2\lambda^{\prime}c_s^2\partial^i\delta\varepsilon.
    \end{align}
    Using the linear order perturbation of the Frenkel condition $S^{\mu\nu}u_{\nu}=0=\omega^{\mu\nu}u_{\nu}$ gives us, $\delta S^{\mu\nu}u_{\nu}^{(0)}=0$. Therefore
due to the use of the Frenkel condition we find $\delta\omega^{0i}=0=\delta S^{0i}$. Using the condition that $\delta S^{0i}=0$ in the above equation we get $\partial^i\delta\varepsilon=0$. Note that for the linear model analysis we are interested in the  plane wave solution of various hydrodynamic perturbations of the form, $e^{-ik\cdot x}$. Thus $\partial^i\delta\varepsilon=0$ implies that $\delta\varepsilon$ itself vanishes, i.e. $\delta \varepsilon=0$ and  $\delta q^i=0$. 
Using Eq.~\eqref{equ19ver1} the evolution equation of $\delta S^{ij}$ can be written as, 
    \begin{align}
\partial_0\delta S^{ij}+2D_s\delta S^{ij}+2\gamma^{\prime}(\partial^i\delta\pi^j-\partial^j\delta\pi^i)=0,
\end{align}
here, $\delta\pi^i=(\varepsilon_{(0)}+p_{(0)})\delta u^i$. 
Now let us look into the conservation of the energy-momentum tensor. 
Using the conditions $\delta\varepsilon=0$ and $\delta q^i=0$ in the longitudinal projection of the energy-momentum tensor, i.e. Eq.~\eqref{equ24ver1} we find, $\partial_i\delta\pi^i=0$. On the other hand 
the perturbation equation associated with the normal projection of conservation of total energy-momentum tensor give us,
\begin{align}
     (\varepsilon_{(0)}+p_{(0)})\partial_0\delta u^{i}+\eta \partial^k\partial_k\delta u^i+\partial_k\delta\phi^{ki}=0.
\label{equ57ver2}
\end{align}
Using the expression of $\delta\phi^{\mu\nu}$ as given in Eq.~\eqref{equ54ver2} it can be shown that, 
\begin{align}
    \partial_k\delta\phi^{ki}= \widetilde{\gamma}\partial_k\partial^k\delta u^i+D_s\partial_k\delta S^{ki}.
\label{equ58ver2}
\end{align}
Using Eq.~\eqref{equ58ver2} back into Eq.~\eqref{equ57ver2} we obtain,
\begin{align}
\partial_0\delta\pi^i+(\gamma_{\perp}+\gamma^{\prime})\partial_k\partial^k\delta \pi^i+D_s\partial_k\delta S^{ki}=0.
\end{align}
In summary, considering the linear order perturbations of various hydrodynamic variables, the conservation of the energy-momentum tensor and the total angular momentum tensor give us
\begin{align}
    &\partial_0\delta\pi^i+(\gamma_{\perp}+\gamma^{\prime})\partial^{k}\partial_k\delta\pi^i+D_s\partial_k\delta S^{ki}=0,\label{equ39ver1}\\
    & \partial_0\delta S^{ij}+2D_s\delta S^{ij}+2\gamma^{\prime}(\partial^i\delta\pi^j-\partial^j\delta\pi^i)=0,\label{equ40ver1}\\
    & \partial_i\delta\pi^i=0. \label{equ41ver1}
\end{align}
Once again without the loss of generality, we consider plane wave representation of various perturbations in the momentum space, 
\begin{align}
   & \delta\pi^k=\widetilde{\delta\pi^k}~e^{-i\omega t+ik_z z}\\
    & \delta S^{ij}=\widetilde{\delta S^{ij}}~e^{-i\omega t+ik_z z}
\end{align}
In the momentum space Eq.~\eqref{equ41ver1} implies $\delta\pi^z=0$. Therefore, the perturbations $\widetilde{\delta\varepsilon}$, $\widetilde{\delta\pi^z}$, $\widetilde{\delta S^{0x}}$, $\widetilde{\delta S^{0y}}$, $\widetilde{\delta S^{0z}}$ decouples from the theory and the remaining non-trivial equations are, 
\begin{align}
& -i\omega\widetilde{\delta\pi^x}+(\gamma_{\perp}+\gamma^{\prime})k_z^2\widetilde{\delta\pi^x}+ik_zD_s\widetilde{\delta S^{zx}}=0\\
& -i\omega\widetilde{\delta\pi^y}+(\gamma_{\perp}+\gamma^{\prime})k_z^2\widetilde{\delta\pi^y}+ik_zD_s\widetilde{\delta S^{zy}}=0\\
& -i\omega\widetilde{\delta S^{xy}}+2D_s\widetilde{\delta S^{xy}}=0\\
& -i\omega\widetilde{\delta S^{zx}}+2D_s \widetilde{\delta S^{zx}}-2\gamma^{\prime}(ik_z)\widetilde{\delta \pi^x}=0\\
& -i\omega\widetilde{\delta S^{yz}}+2D_s \widetilde{\delta S^{yz}}+2\gamma^{\prime}(ik_z)\widetilde{\delta \pi^y}=0
\end{align}
$\widetilde{\delta\pi^x},\widetilde{\delta S^{zx}}$ and $\widetilde{\delta\pi^y},\widetilde{\delta S^{zy}}$ can be grouped together as given in Eq.~\eqref{equ40ver2} and ~\eqref{equ41ver2}. But $\widetilde{\delta S^{xy}}$ does not couple to any other perturbation. Various  dispersion relations can be obtained by solving the following equation, $  \text{det}(\mathcal{B})^2(-i\omega+2D_s)=0$,
and can be summarized as, 
\begin{align}
    &\omega=-2iD_{s}\\
    &\omega=-2iD_{s}-i\gamma^{\prime}k_{z}^{2}+\mathcal{O}(k_{z}^{2})~~~ (\mbox{two modes})\\
    &\omega=-i\gamma_{\perp}k_{z}^{2}+\mathcal{O}(k_{z}^{4})\quad (\mbox{two modes}).
\end{align}
Note that, unlike the previous case, the imaginary part of the various dispersion relation is always negative and various perturbations will not grow with time. But this comes with a drawback. Due to the use of the Frenkel condition in this case, the standard hydrodynamic perturbations, i.e. $\delta\varepsilon$ and $\delta\pi^z$ do not appear in the theory which is not physically appealing.
\section{Spin chemical potential leading order in the hydrodynamic gradient expansion}
\label{sec3}
\subsection{Formulation}
We start our discussion with the hydrodynamic framework with angular momentum where the spin chemical potential ($\omega^{\mu\nu}$) is $\mathcal{O}(1)$ in the hydrodynamic gradient expansion~\cite{She:2021lhe}. In this framework, the energy-momentum tensor is totally symmetric. Note that in the absence of any anti-symmetric component of the energy-momentum tensor, the spin angular momentum is separately conserved even in the presence of interactions which is otherwise not possible. This drastically affects the dissipative spin-hydrodynamic description, particularly various dissipative currents in the energy-momentum tensor and spin tensor~\cite{She:2021lhe}. Here we proceed one step further to discuss spin-hydrodynamic modes of this theory by performing the linear mode analysis for the spin-hydrodynamic framework. We consider the metric convention with $g_{\mu\nu}=\text{diag}(1,-1,-1,-1)$ and projector orthogonal to the fluid four-velocity $u^{\mu}$ is  $\Delta^{\mu\nu}=g^{\mu\nu}-u^{\mu}u^{\nu}$ with $u^{\mu}u_{\mu}=1$. For the Landau frame choice,
the energy-momentum tensor $(T^{\mu\nu})$ and the spin tensor $(S^{\mu\alpha\beta})$ can be expressed as~\cite{She:2021lhe},
\begin{align}
    & T^{\mu\nu}=\varepsilon u^{\mu}u^{\nu}-p\Delta^{\mu\nu}+\tau^{\mu\nu}\\
    & \tau^{\mu\nu}=\pi^{\mu\nu}+\Pi\Delta^{\mu\nu}\\
    & S^{\mu\alpha\beta}=u^{\mu}S^{\alpha\beta}+ S^{\mu\alpha\beta}_{(1)}.
\end{align}
$\tau^{\mu\nu}=\tau^{\nu\mu}$ is the dissipative part of the energy-momentum tensor which contains the shear and bulk viscous terms. $\tau^{\mu\nu}$ satisfies the following condition, $\tau^{\mu\nu}u_{\mu}=0$. Using the entropy current analysis for the Navier-Stokes theory shear and bulk viscous terms can be expressed as, $\tau^{\mu\nu} = \pi^{\mu\nu}+\Pi\Delta^{\mu\nu}$,
\begin{align}
& \pi^{\mu\nu}= \eta\bigg[\nabla^{\mu}u^{\nu}+\nabla^{\nu}u^{\mu}-\frac{2}{3}\Delta^{\mu\nu}\nabla_{\beta}u^{\beta}\bigg],  \\
& \Pi=\zeta (\partial_{\alpha}u^{\alpha}).
\end{align}
Here $\nabla^{\mu}=\Delta^{\mu\nu}\partial_{\nu}$, $\eta$ and $\zeta$ are the coefficients of the shear and bulk viscosity. Both $\eta$ and $\zeta$ are positive definite. The spin tensor $S^{\mu\alpha\beta}$ is only anti-symmetric in last two indices. 
$S^{\mu\nu}$ can be considered as the spin density, i.e., $S^{\alpha\beta}\equiv u_{\mu}S^{\mu\alpha\beta}$, and  $S^{\mu\alpha\beta}_{(1)}$ is the dissipative part of the spin tensor. 
The thermodynamic relations are given as, $Ts+S^{\alpha\beta}\omega_{\alpha\beta}=\varepsilon+p$, $d\varepsilon=Tds+\omega_{\alpha\beta}dS^{\alpha\beta}$. The dissipative part of the spin tensor $ S^{\mu\alpha\beta}_{(1)}\sim \mathcal{O}(\partial) $ can be expressed as \cite{She:2021lhe,Becattini:2011ev}\footnote{In the standard hydrodynamic framework without a dynamical spin degree of freedom the non-equilibrium entropy current can be expressed as, $\mathcal{S}^{\mu}=p\beta u^{\mu}+\beta u_{\nu} T^{\mu \nu}$. In spin hydrodynamics the ansatz for the non-equilibrium entropy current can be generalized to, $\mathcal{S}^{\mu}=p \beta^{\mu}+\beta_{\nu} T^{\mu \nu}-\beta \omega_{\alpha \beta} S^{\mu \alpha \beta}$. If we consider $\omega^{\mu\nu}\sim \mathcal{O}(1)$ then for the first order theory of dissipative spin hydrodynamics $S^{\mu\alpha\beta}_{(1)}$ contributes to the entropy current. }, 
\begin{align}
    & S^{\mu\alpha\beta}_{(1)}= -\frac{\mathfrak{q}^{\mu}}{\varepsilon+p}S^{\alpha\beta}+u^{\alpha}\Delta^{\mu\beta}\Phi-u^{\beta}\Delta^{\mu\alpha}\Phi\nonumber\\
    & ~~~~~~~~~+u^{\alpha}\tau^{\mu\beta}_{(s)}-u^{\beta}\tau^{\mu\alpha}_{(s)}+u^{\alpha}\tau^{\mu\beta}_{(a)}-u^{\beta}\tau^{\mu\alpha}_{(a)}+\Theta^{\mu\alpha\beta}.
    \label{equ5ver3}
\end{align}
Similar to $S^{\mu\alpha\beta}$, the dissipative part $S^{\mu\alpha\beta}_{(1)}$ is also anti-symmetric in last two indices. Various dissipative part in the spin tensor, i.e., $\mathfrak{q}^{\mu}, \Phi,\tau^{\mu\nu}_{(s)}, \tau^{\mu\nu}_{(a)} $, and $\Theta^{\mu\alpha\beta}$ are first order in the hydrodynamic gradient expansion. These dissipative currents satisfy the following properties: $u_{\mu} \mathfrak{q}^{\mu}=u_{\mu} \tau_{(s)}^{\mu \beta}=$ $u_{\mu} \tau_{(a)}^{\mu \beta}=u_{\mu} \Theta^{\mu \alpha \beta}=0 ; \tau_{(s)}^{\mu \beta}=\tau_{(s)}^{\beta \mu}, \tau_{(a)}^{\mu \beta}=$ $-\tau_{(a)}^{\beta \mu}, \Theta^{\mu \alpha \beta}=-\Theta^{\mu \beta \alpha} ; \operatorname{tr}\left(\tau_{(s)}^{\beta \mu}\right)=0$. Note that $\tau^{\mu\nu}_{(s)}$  is a symmetric tensor. Hence in general it can be decomposed into a trace part and a trace-less part.  Trace part of $\tau^{\mu\nu}_{(s)}$ can be absorbed in $\Phi$, hence we consider $\tau^{\mu\nu}_{(s)}$ to be trace-less. Considering that for a dissipative system the entropy will be produced, the analytic expressions for $\mathfrak{q}^{\mu}$, $\Phi$, $\tau^{\mu\nu}_{(s)}$, $\tau^{\mu\nu}_{(a)}$, and $\Theta^{\mu\alpha\beta}$ can be given as~\cite{She:2021lhe}, 
\begin{align}
    & \mathfrak{q}^{\mu}=\lambda_q T \bigg(\frac{\nabla^{\mu}T}{T}-Du^{\mu}\bigg), \\
    & \Phi = -\chi_1 u^{\alpha}\nabla^{\beta}(\beta\omega_{\alpha\beta}), \\
    &     \tau^{\mu\beta}_{(s)}  =-\chi_2u^{\alpha}\bigg[\Delta^{\beta\rho}\Delta^{\mu\gamma}+\Delta^{\mu\rho}\Delta^{\beta\gamma}\nonumber\\
    & ~~~~~~~~~~~~~~~~~~~~~~~~~~-\frac{2}{3}\Delta^{\mu\beta}\Delta^{\rho\gamma}\bigg]\nabla_{\gamma}\bigg(\beta \omega_{\alpha\rho}\bigg),\\
    & 
\tau_{(a)}^{\mu \beta}=-\chi_{3} u^{\alpha}\left(\Delta^{\beta \rho} \Delta^{\mu \gamma}-\Delta^{\mu \rho} \Delta^{\beta \gamma}\right) \nabla_{\gamma}\left(\beta\omega_{\alpha \rho}\right),\\
&     \Theta^{\mu\alpha\beta}=-\chi_4\bigg[u^{\beta}u^{\rho}\Delta^{\alpha\delta}-u^{\alpha}u^{\rho}\Delta^{\beta\delta}\bigg]\Delta^{\mu\gamma}\nabla_{\gamma}(\beta\omega_{\delta\rho})\nonumber\\
& ~~~~~~~~~~~~~+\chi_5\Delta^{\alpha\delta}\Delta^{\beta\rho}\Delta^{\mu\gamma}\nabla_{\gamma}(\beta\omega_{\delta\rho}),
\end{align}
Here $D\equiv u^{\mu}\partial_{\mu}$. Various transport coefficients, $\lambda_q, \chi_1,\chi_2,\chi_3,\chi_4$, and $\chi_5$ are all positive. We should emphasize the relative sign difference between two terms in the expression of $\Theta^{\mu\alpha\beta}$. Although $\Theta^{\mu\alpha\beta}$ is orthogonal to the fluid four-velocity, in the fluid rest frame it is not totally space-like. In the fluid rest frame terms associated with $\chi_5$ have only space-like indices, but the terms associated with $\chi_4$ can have space-like and time-like indices. This gives rise to a relative sign difference.  

To study the linear mode analysis of the spin hydrodynamic description we consider the global equilibrium background flow $u^{\mu}_{(0)}\equiv(1,0,0,0)$, with  $S^{\mu\nu}_{(0)}=0$ and $\omega^{\mu\nu}_{(0)}=0$. 
All other dissipative currents vanish in the global equilibrium background. Fluid flow perturbation is given as $\delta u^{\mu}=(0,\delta u^i)$. Taking the projection of the conservation of $T^{\mu\nu}$ along with the fluid velocity and orthogonal to the fluid velocity we find, 
\begin{align}
    & u^{\mu}\partial_{\mu}\varepsilon+(\varepsilon+p)(\partial_{\mu}u^{\mu})=-u_{\nu}\partial_{\mu}\tau^{\mu\nu},\\
    &  (\varepsilon+p)Du^{\alpha}-\Delta^{\alpha\beta}\partial_{\beta}p+\Delta^{\alpha}_{~\nu}\partial_{\mu}\tau^{\mu\nu}=0.
\end{align}
Corresponding linear order perturbation equations can be expressed as, 
\begin{align}
    & \partial_{0}\delta\varepsilon+\partial_i\delta \pi^{i}=0,\label{equ13ver3} \\
    & \partial_0\delta \pi^{i}-c_s^2\partial^i\delta\varepsilon+\gamma_{\perp}\bigg[ \delta^i_l\partial^k\partial_k- \partial^i\partial_l\bigg]\delta \pi^l +\gamma_{||} \partial^i\partial_k\delta \pi^k = 0.\label{equ14ver3}
\end{align}
Here, $\delta\pi^i=(\varepsilon_{(0)}+p_{(0)})\delta u^i$, $\gamma_{\perp}\equiv \eta/(\varepsilon_{(0)}+p_{(0)})$ and $\gamma_{||}\equiv (\frac{4}{3}\eta+\zeta)/(\varepsilon_{(0)}+p_{(0)})$, $c_s^2=\partial p/\partial \varepsilon$. 
$\varepsilon_{(0)}$, $p_{(0)}$ are the background energy density and pressure. To obtain the above perturbation equations we have used the following equation,  
\begin{align}
    \delta\tau^{\mu\nu} & =\eta\bigg[\Delta^{\mu\alpha}_{(0)}\partial_{\alpha}\delta u^{\nu}+\Delta^{\nu\alpha}_{(0)}\partial_{\alpha}\delta u^{\mu}-\frac{2}{3}\Delta^{\mu\nu}_{(0)}\partial_{\alpha}\delta u^{\alpha}\bigg]\nonumber\\
    &~~~ +\zeta \Delta^{\mu\nu}_{(0)}\partial_{\alpha}\delta u^{\alpha}+\mathcal{O}(\partial^3).
\end{align}
It should be emphasized that for the background flow with $u^{\mu}_{(0)}\equiv (1,0,0,0)$, $\delta\tau^{00}=0$ as well as $\delta\tau^{i0}=0$. 
Since the energy-momentum tensor does not have any anti-symmetric part, the spin tensor $S^{\mu\alpha\beta}$ is separately conserved. This follows from the conservation of the total angular momentum, i.e.,
\begin{align}
& u^{\mu}\partial_{\mu}S^{\alpha\beta}+S^{\alpha\beta}(\partial_{\mu}u^{\mu})+\partial_{\mu}S^{\mu\alpha\beta}_{(1)}=0. 
\end{align}
For the global equilibrium condition with $S^{\alpha\beta}_{(0)}=0$, the perturbation equation corresponding to the conservation of the spin tensor can be expressed as,
\begin{align}
u^{\mu}_{(0)}\partial_{\mu}\delta S^{\alpha\beta}+\partial_{\mu}\delta S^{\mu\alpha\beta}_{(1)}=0.
\label{equ26ver1}
\end{align}
In the above equation $\delta S^{\mu\alpha\beta}_{(1)}$ can be obtained systematically by taking linear order perturbation of Eq.~\eqref{equ5ver3} for the given global equilibrium conditions.  To obtain $\delta S^{\mu\alpha\beta}_{(1)}$ we use the following relations, 
\begin{align}
    & \delta\Phi = -\chi_1\beta_{(0)}\partial_i\delta \omega^{0i}, \\
    &  \delta \tau^{\mu\alpha}_s  = -\beta_{(0)} \chi_2u^{a}_{(0)}\bigg[\Delta^{\alpha\rho}_{(0)}\Delta^{\mu\gamma}_{(0)}+\Delta^{\mu\rho}_{(0)}\Delta^{\alpha\gamma}_{(0)}\nonumber\\
    & ~~~~~~~~~~~~~~~~~~~~~~~~~-\frac{2}{3}\Delta^{\mu\alpha}_{(0)}\Delta^{\rho\gamma}_{(0)}\bigg]\nabla_{\gamma}^{(0)} \delta\omega_{a\rho},\\
    & \delta\tau_{(a)}^{\mu \beta}=-\beta_{(0)}\chi_{3} u^{\alpha}_{(0)}\left(\Delta^{\beta \rho}_{(0)} \Delta^{\mu \gamma}_{(0)}-\Delta^{\mu \rho}_{(0)} \Delta^{\beta \gamma}_{(0)}\right) \nabla_{\gamma}^{(0)}\delta\omega_{\alpha \rho},\\
    & \delta \Theta^{\mu\alpha\beta}=-\chi_4\beta_{(0)}\bigg[u^{\beta}_{(0)}u^{\rho}_{(0)}\Delta^{\alpha\delta}_{(0)}-\nonumber\\
    & ~~~~~~~~~~~~~~~~~~~~~~~~~~~~~~~~u^{\alpha}_{(0)}u^{\rho}_{(0)}\Delta^{\beta\delta}_{(0)}\bigg]\Delta^{\mu\gamma}_{(0)}\nabla_{\gamma}^{(0)}\delta\omega_{\delta\rho}\nonumber\\
    & ~~~~~~~~~~~~~~~~~~~~~~+\chi_5\beta_{(0)}\Delta^{\alpha\delta}_{(0)}\Delta^{\beta\rho}_{(0)}\Delta^{\mu\gamma}_{(0)}\nabla_{\gamma}^{(0)}~\delta\omega_{\delta\rho}.
\end{align}
In Eq.~\eqref{equ5ver3} $\mathfrak{q}^{\mu}$ appears with $S^{\alpha\beta}$.
Since for the global equilibrium $\mathfrak{q}^{\mu}$ as well as $S^{\alpha\beta}$ vanishes, in $\delta S^{\mu\alpha\beta}_{(1)}$ , $\delta \mathfrak{q}^{\mu}$ does not contribute.
The `$0i$'-th component of Eq.~\eqref{equ26ver1} can be expressed as,
\begin{align}
   & \widetilde{\chi}_b\partial_0\delta\omega^{0i}-\chi_1\beta_{(0)}\partial^i\partial_l\delta \omega^{0l}\nonumber\\
   & -\beta_{(0)}\chi_2\bigg[\partial_j\partial^j\delta\omega^{0i}+\partial_j\partial^i\delta\omega^{0j}-\frac{2}{3}\partial^i\partial_{k}\delta\omega^{0k}\bigg]\nonumber\\
    &-\beta_{(0)} \chi_3 (\partial_j\partial^j\delta\omega^{0i}-\partial_j\partial^i\delta\omega^{0j})-\beta_{0}\chi_4 \partial_j\partial^j\delta\omega^{0i}=0,
\label{equ18ver3}
\end{align}
and the `$ij$'-th component of Eq.~\eqref{equ26ver1} can be expressed as, 
\begin{align}
\widetilde{\chi}_s \partial_0\delta \omega^{ij}+\beta_{(0)}\chi_5\partial_l\partial^l\delta\omega^{ij}=0.   
\label{equ19ver3}
\end{align}
In Eqs.~\eqref{equ18ver3} and \eqref{equ19ver3} we define,  $\widetilde{\chi}_b\equiv \partial S^{0i}/\partial\omega^{0i}$ and $\widetilde{\chi}_s\equiv \partial S^{ij}/\partial\omega^{ij}$. 
$\beta_{(0)}$ denotes the inverse  global equilibrium temperature. 
Eqs.~\eqref{equ13ver3}, \eqref{equ14ver3}, \eqref{equ18ver3}, and \eqref{equ19ver3}
are the perturbation equations which we can analyze in the momentum space.  From these equations one may observe that perturbation in the standard hydrodynamic variables, i.e., $\delta\varepsilon$ and $\delta\pi^i$ are decoupled from the perturbation in the spin degree of freedom, i.e., $\delta S^{0i}$, $\delta S^{ij}$ or $\delta \omega^{0i}$, $\delta \omega^{ij}$. This is the artifact of the following factors, (i) in this framework the energy-momentum tensor is symmetric, (ii) the spin chemical potential and spin density are both $\mathcal{O}(1)$ in the hydrodynamic gradient expansion, (iii) we consider the global equilibrium with $S^{\mu\nu}_{(0)}=0$ and $\omega^{\mu\nu}_{(0)}=0$.

\subsection{Fourier space equations}
We look for the plane wave solution of the form $e^{-i\omega t+i \vec{k}\cdot\vec{ x}}$. Various perturbations, $\delta\varepsilon$, $\delta\pi^k$, $\delta \omega^{ij}$ and $\delta \omega^{0i}$ can be expressed as plane waves in the following way \begin{align}
    & \delta\varepsilon=\widetilde{\delta\varepsilon}~e^{-i\omega t+i\vec{k}\cdot\vec{x}}\nonumber\\
    & \delta\pi^k=\widetilde{\delta\pi^k}~e^{-i\omega t+i\vec{k}\cdot\vec{x}}\nonumber\\
    & \delta \omega^{ij}=\widetilde{\delta \omega^{ij}}~e^{-i\omega t+i\vec{k}\cdot\vec{x}}\nonumber\\
    & \delta \omega^{0i}=\widetilde{\delta \omega^{0i}}~e^{-i\omega t+i\vec{k}\cdot\vec{x}}.
\end{align}
Again for the rotational symmetry of the system one may consider plane wave along the z-direction, i.e. $\vec{k}\equiv(0,0,k_z)$~\cite{Hattori:2019lfp}. For such a waveform, 
Eqs.~\eqref{equ13ver3}, \eqref{equ14ver3}, \eqref{equ18ver3}, and \eqref{equ19ver3} can be expressed as, 
\begin{align}
    & -i\omega~\widetilde{\delta \varepsilon}+i k_z~\widetilde{\delta\pi^z}=0,\label{equ20ver3}\\
    & -i\omega~\widetilde{\delta\pi^x} +\gamma_{\perp}k_z^2~\widetilde{\delta\pi^x}=0,\label{equ21ver3}\\
      & -i\omega~\widetilde{\delta\pi^y} +\gamma_{\perp}k_z^2~\widetilde{\delta\pi^y}=0,\label{equ22ver3}\\
      & -i\omega~\widetilde{\delta\pi^z}+c_s^2 ik_z~ \widetilde{\delta \varepsilon}+\gamma_{||}k_z^2~ \widetilde{\delta\pi^z} = 0,\label{equ23ver3}\\
    & -i\omega~\widetilde{\chi}_b\delta\widetilde{\omega^{0x}}-\beta_{(0)}k_z^2(\chi_2+\chi_3+\chi_4)\delta\widetilde{\omega^{0x}}=0,\label{equ24ver3}\\
    & -i\omega~\widetilde{\chi}_b\delta\widetilde{\omega^{0y}}-\beta_{(0)}k_z^2(\chi_2+\chi_3+\chi_4)\delta\widetilde{\omega^{0y}}=0, \label{equ25ver3}\\
   & -i\omega~\widetilde{\chi}_b\delta\widetilde{\omega^{0z}}-\beta_{(0)}k_z^2(\chi_1+\frac{4}{3}\chi_2+\chi_4)\delta\widetilde{\omega^{0z}}=0, \label{equ26ver3}\\
   & -i\omega~\widetilde{\chi}_s \widetilde{\delta\omega^{kl}}+\chi_5\beta_{0} k_z^2 \widetilde{\delta\omega^{kl}} = 0.\label{equ27ver3} 
   \end{align}
   Eqs.~\eqref{equ20ver3}-\eqref{equ23ver3} give us the standard hydrodynamic modes:
   \begin{align}
       & \omega =-i\gamma_{\perp}k_z^2.\quad\mbox{(two modes)}\\
       & \omega = \pm c_s k_z-\frac{i}{2}\gamma_{||}k_z^2+\mathcal{O}(k_z^3).
   \end{align}
   Similarly the 
   the spin-wave mode associated with $\delta \omega^{0x}$ and $\delta \omega^{0y}$ are same and can be given as, 
   \begin{align}
       \omega = i\beta_{(0)}k_z^2 \left(\frac{\chi_2+\chi_3+\chi_4}{\widetilde{\chi}_b}\right).
   \end{align}
   The wave mode associated with $\delta \omega^{0z}$ can be expressed as, 
   \begin{align}
       \omega = i\beta_{(0)}k_z^2 \left(\frac{\chi_1+\frac{4}{3}\chi_2+\chi_4}{\widetilde{\chi}_b}\right).
   \end{align}
   Finally, the wave modes associated with $\delta \omega^{kl}$:
   \begin{align}
       \omega=-i\frac{\chi_5}{\widetilde{\chi}_s}\beta_{0} k_z^2.
   \end{align}
   Note that $\chi_1$, $\chi_2$, $\chi_3$, $\chi_4$, $\chi_5$, $\widetilde{\chi}_b$ and $\widetilde{\chi}_s$ are all considered to be positive. Wave modes associated with $\omega^{0i}$ or $S^{0i}$ have a positive imaginary part. Therefore these modes are not stable and the corresponding modes will not decay in time. Such growing modes are not physically appealing. Such modes can be removed if we use the Frenkel condition with $S^{\mu\nu}u_{\nu}=0$. At the linear order perturbation level, for the global equilibrium with $S^{\mu\nu}_{(0)}$ the Frenkel condition gives us $\delta S^{\mu\nu}u_{\nu}^{(0)}=0$. This implies $\delta S^{0i}=0$ for the background flow with $u^{\mu}=(1,0,0,0)$. Interestingly if we impose the Frenkel condition then the problematic spin-modes will no longer appear in the calculation. Moreover, this does not affect any standard hydrodynamic modes as the standard hydrodynamic mode and the spin modes are decoupled in this framework.  
\section{Summary}
\label{sec4}
We examine the solutions of the spin-hydrodynamic equations in Fourier space for two different first-order spin-hydrodynamic formulations. The first considers the spin chemical potential $\omega_{\mu\nu}$ to be first order in gradient expansion \cite{Hattori:2019lfp}, while the other is of leading order \cite{She:2021lhe}. Our calculation suggests that unstable solutions for the spin $S^{0i}$ component may emerge from spin hydrodynamic equations.
Frenkel condition can be used to get rid of such generic instabilities. But when we consider the Frenkel condition along with $\omega^{\mu\nu}\sim\mathcal{O}(\partial)$ then the standard hydrodynamic modes also get affected. Such an unwanted feature is absent when we consider the case with $\omega^{\mu\nu}\sim\mathcal{O}(1)$. The stability of spin modes is not a settled issue in literature and some physical understanding may be required to construct a proper model of first-order spin hydrodynamics. One such possibility is to look into second-order spin-hydrodynamic formulation unless the first-order spin-hydrodynamic formulation can be shown to be stable for generic configurations. In the context of standard hydrodynamic theory it has been argued that instability can arise due to acausal modes. In this calculation we have not studied the causality of linear modes. Moreover the stability property has been considered at the linear level for un-polarized background. It will be interesting to investigate the stability property for a generic polarized background.

\medskip
{\it Acknowledgements.}  This work was supported in part by the Polish National Science Centre Grant No. 2018/30/E/ST2/00432. We would like to thank Duan She, Koichi Hattori, Wojciech Florkowski, and Victor Ambrus for important discussions on the spin hydrodynamic frameworks.

\bibliography{ref.bib}{}
\bibliographystyle{utphys}

\end{document}